\documentclass[twocolumn,10pt,final,conference]{IEEEtran}
\makeatletter
\def\ps@headings{%
\def\@oddhead{\mbox{}\scriptsize\rightmark \hfil \thepage}%
\def\@evenhead{\scriptsize\thepage \hfil\leftmark\mbox{}}%
\def\@oddfoot{}%
\def\@evenfoot{}}
\makeatother 
\usepackage{supertabular}
\usepackage{cite}
\usepackage{amsfonts}
\usepackage[cmex10]{amsmath}
\usepackage{array}
\usepackage{mdwmath}
\usepackage{eqparbox}
\usepackage{graphics}
\usepackage{graphicx}
\usepackage{epsfig}
\usepackage{subfigure}
\usepackage{subfig}
\usepackage{amssymb}
\usepackage{lineno}
\usepackage{cite}
\usepackage{booktabs}
\usepackage{tabularx}
\usepackage{amsmath}
\usepackage{multirow}
\usepackage{epsfig}
\usepackage{amssymb}
\usepackage{algorithmic}
\usepackage[ruled]{algorithm}
\usepackage{color}
\usepackage{times}
\usepackage{paralist}
\usepackage{xspace}
\usepackage{mathrsfs}
\usepackage{clrscode}
\usepackage{epstopdf}
\usepackage{caption}
\allowdisplaybreaks 

\newtheorem{defn}{Definition}

\renewcommand{\paragraph}[1]{\smallskip \noindent {\textsc{#1}}}

\renewcommand{\baselinestretch}{1}
\usepackage{geometry}
\def\ie{\textit{i.e.}\xspace}
\def\etal{\textit{et al.}\xspace}
\def\etc{\textit{etc.}\xspace}
\def\eg{\textit{e.g.}\xspace}

\begin{document}

\title{STC: Coarse-Grained Vehicular Data Based Travel Speed Sensing by Leveraging Spatial-Temporal Correlation}
\author{
\IEEEauthorblockN{Lu Shao, Cheng Wang, Changjun Jiang}
\IEEEauthorblockA{ Department of Computer Science, Tongji University, Shanghai, China, 201804}
}
\maketitle
\thispagestyle{empty}

\begin{abstract}
As an important information for traffic condition evaluation, trip planning, transportation management, \etc, average travel speed for a road means the average speed of vehicles travelling through this road in a given time duration.
Traditional ways for collecting travel-speed oriented traffic data always depend on dedicated sensors and supporting infrastructures, and are therefore financial costly.
Differently, vehicular crowdsensing as an infrastructure-free way, can be used to collect data including real-time locations and velocities of vehicles for road travel speed estimation, which is a quite low-cost way. However, vehicular crowdsensing data is always coarse-grained. This coarseness can lead to the incompleteness of travel speeds. Aiming to handle this problem as well as estimate travel speed accurately, in this paper, we propose an approach named STC that exploits the \emph{spatial-temporal correlation} among travel speeds for roads by introducing the time-lagged cross correlation function. The \emph{time lagging factor} describes the time consumption of traffic feature diffusion along roads. To properly calculate cross correlation, we novelly make the determination of the time lagging factor self-adaptive by recording the locations of vehicles at different roads. Then, utilizing the local stationarity of cross correlation, we further reduce the problem of single-road travel speed vacancy completion to a minimization problem. Finally, we fill all the vacancies of travel speed for roads in a recursive way using the geometric structure of road net. 
Elaborate experiments based on real taxi trace data show that STC can settle the incompleteness problem of vehicle crowdsensing data based travel speed estimation and ensure the accuracy of estimated travel speed better, in comparison with representative existing methods such as KNN, Kriging and ARIMA.
\end{abstract}

\section{Introduction}\label{section-Introduction}
Average travel speed for a road means the average speed of vehicles travelling through this road in a given time duration.
Such information can be used for traffic condition evaluation, trip planning, transportation management, and so on.
Providing services based on average travel speed is very helpful both for drivers and traffic controllers.
To get travel speed information, in traditional ways, the road net is monitored by infrastructure-supported devices such as cameras \cite{Ref72,Ref73}, loop detectors \cite{Ref74,Ref75} and radio sensors \cite{Ref71}  which are very expensive to deploy and maintain \cite{Ref79}.
What's more, these ways can be affected by environment. For example, quality of camera-based data is vulnerable to weather condition, light condition, and so on.

Different from infrastructure-based ways, vehicular crowdsensing is an up-to-the-moment infrastructure-free way for traffic data collection.
Vehicles running on roads can incidentally offer their data including timestamps, locations and real-time speeds to a central server via wireless communication.
The feasibility of vehicular crowdsensing is guaranteed by the wide deployment of on-board GPS and wireless communication devices such as navigation facilities. Since it almost needs no additional equipment, the cost of vehicular crowdsensing is greatly low when compared with infrastructure-based methods. We can use vehicular crowdsensing data to estimate travel speeds for roads.
However, vehicular crowdsensing data is usually coarse-grained. The first reason is that not all the vehicles running on roads will be willing to offer data. The second reason is the temporal variation of spatial distribution of vehicles. Take it as an example, in morning rush hour, most roads can be covered by vehicles while much less roads can be covered in midnight. The coarseness of vehicular crowdsensing data will lead to the incompleteness problem of travel speeds of roads,  which can be testified by analysis of real taxi probe data. In addition, to guarantee the usability of travel speeds, the estimated values have to meet predefined accuracy requirement.

In order to handle the incompleteness problem of vehicular crowdsensing data based travel speed estimation and to calculate travel speed accurately, in this paper, we propose a spatial-temporal correlation based approach named STC (\emph{\bf{S}}patial \emph{\bf{T}}emporal \emph{\bf{C}}orrelation). In STC, we leverage the spatial-temporal correlation of travel speeds among roads. Our basic argument is that the  travel speeds of roads close to each other are correlated due to spatial proximity of these roads. More specifically, we use the time-lagged cross correlation function to measure the relevances between travel speeds of different roads for different calculation intervals. Note that due to the time consumption for vehicles to run from road $r_{1}$ to $r_{2}$, within a relative small area, the farther the distance between two roads is, the longer the time taken by a traffic characteristic to spread from $r_{1}$ to $r_{2}$ will be. We call this time consumption as \emph{time lagging factor}. Particularly, we employ the traceability of vehicle locations to determine the time lagging factor self-adaptively, which is helpful to properly calculate the cross correlation between roads. Having the cross correlation, we fill the vacancy of single-road travel speed with the help of the near-by roads whose travel speeds are not empty. We accomplish this by converting it to a minimization problem, using the local stationarity of cross correlation between roads. Finally, we fill all the travel speed vacancies of roads in a recursive way by utilizing the geometric structure of road net.

To examine the effectiveness of our approach, we conduct experiments based on real taxi trace data including 8363 taxis, collected in Shenzhen, China during 18-26, April, 2011. Through cross-validation, we show that our approach can handle the incompleteness of travel speeds and guarantee the accuracy of travel speed calculation better, compared with classical estimation methods such as KNN, Kriging and ARIMA.

The main contributions of this paper are as follows:

$\bullet$We propose a spatial-temporal correlation based approach named STC to estimate travel speed based on coarse-grained vehicular crowdsensing data, in which we use time-lagged cross correlation function to measure the correlation among roads in terms of travel speeds.

$\bullet$To properly calculate cross correlation, we novelly  make the determination of time lagging factor self-adaptive by leveraging the traceability of vehicle locations at upstream and downstream roads.

$\bullet$We reduce the problem of single-road travel speed vacancy completion to a minimization problem by using the local-stationarity of cross correlation. Then by utilizing the topology of road net, we use a recursive way to fill all the travel speed vacancies of roads.

The rest of this paper is organized as follows. In Section \ref{section-Related Works}, we make a review of existing works that related to ours. In Section \ref{section-Problem Formulation}, we formulate the problem that we aim to handle.
Then in Section \ref{section-STC} we introduce our design of STC for travel speed estimation in detail.
Next, we show our experimental evaluation results and give a discussion of the use of our method for future travel speed prediction in Section \ref{section-Experimental Evaluation}. Finally, our conclusion and future work will be discussed in Section \ref{section-Conclusion And Future Work}.
\section{Related Work}\label{section-Related Works}
We review  existing works related to ours in following aspects:

\emph{Vehicular Crowdsensing.}
There are works that examined the use of vehicles for certain applications. For example, Bauza \etal \cite{Ref13} proposed a vehicular multi-hop end-to-front local traffic congestion level detection strategy. Du \etal \cite{Ref16} aimed to effectively monitor congestion by floating cars whose routes are optimized to sense data in a wide geographical range. Hu \etal \cite{Ref101} designed a role-transferring mechanism for on-board phones to optimize communication performance in vehicle-based participatory sensing applications. Thiagarajan \etal \cite{VTrack} proposed to track mobile phone locations to estimate road traffic delay, which was an energy-efficient way. Sun \etal \cite{EnablingEmergency} used cognitive radio vehicular networks to enable emergency communication between vehicles. Coric \etal \cite{Ref42} used crowdsensing to sketch the map of on-street parking spaces to help drivers to find parking positions. Chen \etal \cite{Ref76} took privacy into consideration while used participatory sensing to draw high quality city map. None of above works inspected travel speed estimation while it was done by Wang \etal \cite{Ref80} who only paid attention to freeways, rather than ordinary roads. And Aslam \etal \cite{Ref79} used taxis as probes to estimate traffic volume with the help of logistic regression and linear regression. But none of above existing works applied vehicular crowdsensing into the wide range travel speed calculation.

\emph{Vacancy Completion Methods.}
The incompletion problem is a critical problem in vehicle-collected data. To handle this problem, different works in the literature used various methods. For example, Alasmary \etal \cite{Ref41} used branch and bound approximation algorithm to select optimal number of vehicles for collecting data when wireless channel capacity is limited. And compressive sensing was considered by works such as \cite{Ref49} \cite{Ref50} \cite{Ref51} to collect and recover vehicle-related data. However, the computational complexity of compressive sensing is always high.
Matrix completion  was also used as an incomplete data estimation method in the work of Du \etal \cite{Ref16}, which was also of high calculation cost. What's more, we mainly focus on single-calculation-interval vacancies completion which is more like a single-column vector, rather than a multi-column matrix. Taking spatial correlation of roads into consideration, Zou \etal \cite{Ref82} proposed to use Kriging interpolation to complete the data where no detection equipment such as loop detector is deployed. Correlation among roads was better used by Aslam \etal \cite{Ref79} who introduced the similarity of roads. However, the temporal correlation among roads was overlooked by them.

\section{Problem Formulation}\label{section-Problem Formulation}
In this section, we first make clear the formalization of road net, based on which we state the problem of coverage incompletion according to taxi trace statistics. Then we explain the goal of travel speed calculation.

\subsection{Road Net Formalization}
\begin{figure}[h]
\centering
\subfigure[A sketch of road structure.]
{
\includegraphics[width=1.4in]{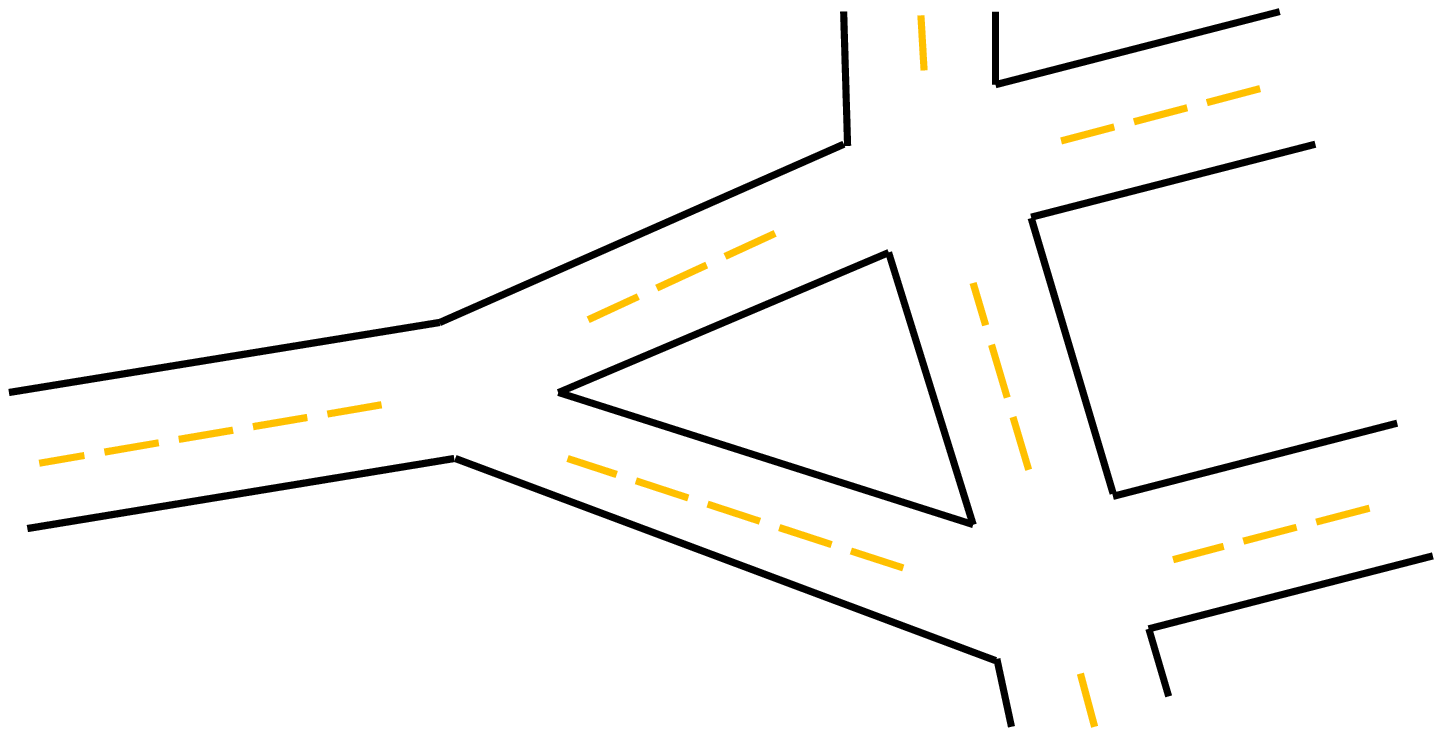}
}
\subfigure[Graphical presentation of road structure.]
{
\includegraphics[width=1.4in]{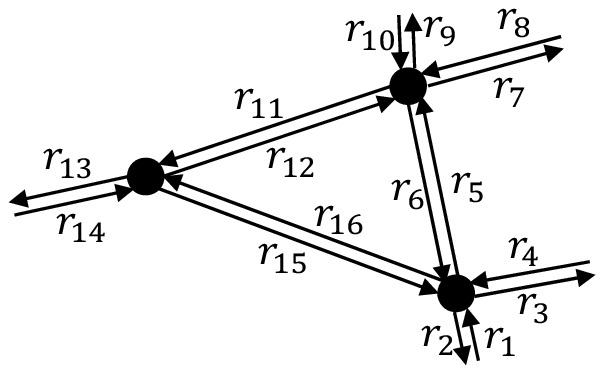}
}
\caption{Illustration of road net formalization.}
\label{fig-Illstration of road net}
\end{figure}

Fig.\ref{fig-Illstration of road net}(a) is a sketch of a fraction of real road net. It consists of \emph{road segments} and \emph{intersections},
A road segment is a portion of road that has only one entrance and only one exit and vehicles are only allowed to run from this entrance to this exit under transportation regulations.
An intersection is a point connecting adjacent road segments. 
For conciseness, we use directed graph to present the road net.  A vertice in this graph represents an intersection in real road net while an edge represents a road segment. We use $\mathcal{G}$ to represent the graph of the whole road net, $\mathcal{R}$ to represent the edge set of $\mathcal{G}$, and $\mathcal{V}$ to represent its vertice set. Fig.\ref{fig-Illstration of road net}(b) is the graphical form of Fig.\ref{fig-Illstration of road net}(a), where $r_i\in \mathcal{R}, i=1,2,\cdots,16$.

\subsection{Problem of Coverage Incompletion in Vehicular Crowdsensing}\label{section-sub-Problem of coverage incompletion in vecicular crowdsensing}
\begin{figure}[h]
  \centering
  \includegraphics[width=\columnwidth]{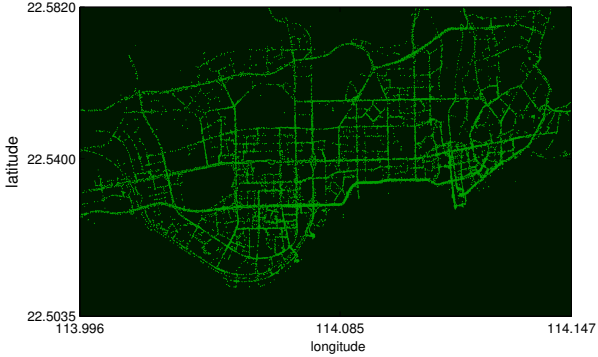}
  \caption{A snapshot of taxi distribution.}
  \label{fig-SnapShot of Taxi Distribution}
\end{figure}

As aforementioned, we use vehicular crowdsensing to collect traffic data. The data containing \emph{timestamp, longitude, latitude, velocity} of a vehicle will be offered at will by this vehicle via wireless communication. We formally call a piece of data in this form as a \emph{record}. We examine real trace data of taxis in Shenzhen, China to first show the spatial gain of our approach and then show the still-existing spatial-temporal vacancies. There are 8363 taxis that participated in this work of data collection from April 18th 00:00:00, 2011 to April 26th 00:00:00, 2011.
Fig.\ref{fig-SnapShot of Taxi Distribution} illustrates a snapshot of the spatial distribution of these taxis. Intuitively, we can see that the roads reached by taxis are not limited to main roads.

However, given the length of calculation interval $T$, owing to the mobility of vehicles, the number of collected records that can be used to directly calculate the travel speeds for different road segments will vary in different intervals. For $i^{th}$ road segment $r_i\in\mathcal{R}$, in $j^{th}$ calculation interval $t_j$, we denote this number as $n_{ij}$.
For trust consideration, only when $n_{ij}$ is equal or bigger than a predefined threshold integer $N_{thr}(T)$, we can calculate the travel speed for $r_i$ directly.
So we define,
\begin{defn}\label{defn-Covered}
Road segment $r_i$ is \emph{covered} in $j^{th} $ calculation interval $t_j$ if $n_{ij}\geqslant N_{thr}(T)$.
\end{defn}

\begin{figure}[h]
  \centering
  \includegraphics[width=\columnwidth]{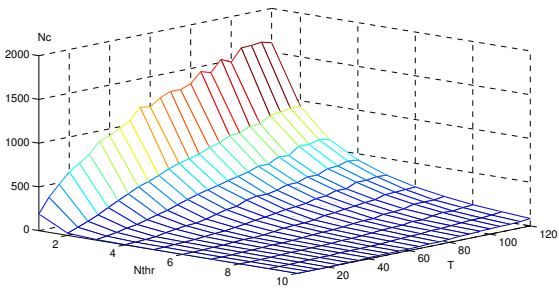}
  \caption{The variation of $N_c$ when $T$ changes from 10s to 120s with step 10s and $N_{thr}$ changes from 1 to 10 with step 1. The begin time of this duration is fixed at 8:00:00 April 18th, 2011.}
  \label{fig-Nc Variation}
\end{figure}

Let $N_R$ denote the total number of road segments in $\mathcal{R}$. Based on our road net formalization, we have 1882 road segments in the  central area of Shenzhen, which gives $N_R=1882$. Then we denote the number of road segments that are covered  in $t_j$ as $N_c$. Fig.\ref{fig-Nc Variation} shows the variation of $N_c$ with different $T$ and $N_{thr}$.

\begin{figure}[h]
  \centering
  \includegraphics[width=\columnwidth]{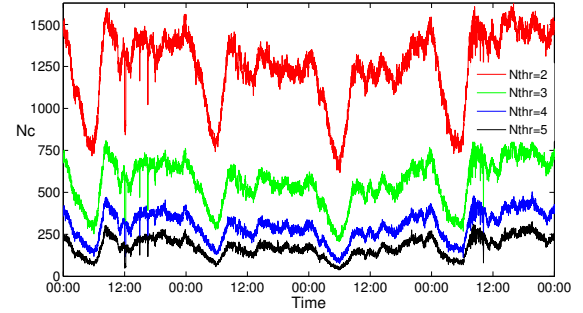}
  \caption{The variation of $N_c$ with the advance of time at fixed $T=60s$ and $N_{thr}=2,3,4,5$ respectively. The time duration is from April 22th 00:00:00, 2011 to April 26th 00:00:00, 2011.}
  \label{fig-Nj When Fix N_thr and T}
\end{figure}

From Fig.\ref{fig-Nc Variation}, we can see that when the value of $T$ is bigger and the value of $N_{thr}$ is smaller, $N_c$ becomes bigger. This trend is consistent with our expectation.
Then we fix $T$, and to see the variation of $N_c$ following the advance of time with different values of $N_{thr}$. The result is shown in Fig.\ref{fig-Nj When Fix N_thr and T}. As we can see, when $N_{thr}$ increases, the value of $N_c$ decreases. Also, the value of $N_c$ has peaks and valleys. These phenomena show us the coverage incompletion of vehicular crowdsensing data.

%

\section{Our Design of STC}\label{section-STC}
In this section, we first give an overview of STC. Then we introduce our strategy for GPS map-matching. Next we explain our methods for single road travel speed calculation and global travel speed vacancy completion.
\subsection{Overview of STC}\label{section-Overview}
\begin{figure}[h]
  \centering
  \includegraphics[width=\columnwidth]{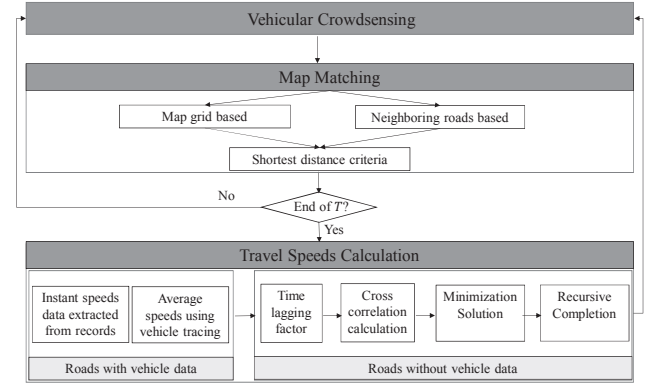}
  \caption{An overview of STC.}
  \label{fig-Overview of architecture}

\end{figure}
The overview of STC is shown in Fig.\ref{fig-Overview of architecture}.
In STC, every time a record is uploaded by vehicular crowdsensing, the GPS information will be extracted from the record and mapped to its corresponding road on map, where we use grid index of roads and vehicle tracking to accelerate map-matching.
When a calculation interval ends, travel speeds of roads where vehicle data is available are calculated by combing instant vehicle speed that is extracted from uploaded records and vehicle tracking based average speeds. Coverage vacancies arise after this, so for the roads without vehicle data, spatial-temporal correlation among travel speeds is considered to fill up these vacancies, in which we make the determination of time lagging factor in cross correlation calculation self-adaptive. Then, we reduce the single-road travel speed vacancy completion problem to a minimization problem using the local-stationarity of cross correlation. Finally, we fulfill global coverage completion in a recursive way.
\subsection{ Map-Matching}\label{section-Map Matching}
The goal of map-matching is to project a GPS location to a road segment accurately. There are two problems we have to handle. The first is the noise contained in GPS data, outliers for example. The second is that there is a large amount of roads as the candidates of a GPS location at the beginning of matching.

\begin{figure}[h]
\centering
\subfigure[Grid split of map.]
{
\includegraphics[width=2in]{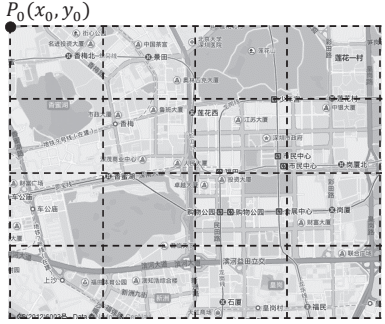}
}
\subfigure[Vehicle tracing based candidate roads selection.]
{
\includegraphics[width=2in]{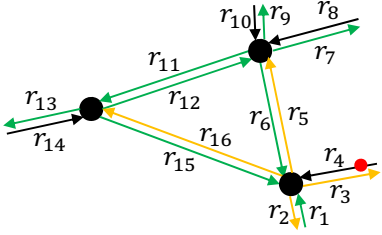}
}
\caption{Grid and vehicle tracing based map matching.}
\label{fig-Map Matching}
\end{figure}
To eliminate the noise in GPS data, we use a constrained shortest distance to find the fittest road segment for $P$. We give a minimum distance $D_{min}$. If the distances between $P$ and any segments are bigger than $D_{min}$, we reckon that $P$ is an outlier and we will drop it.

To handle the candidate choosing  problem, we first split the map into grids and index road segments to the grids.
An illustration of this split is shown in Fig.\ref{fig-Map Matching}(a).
Note that there are many roads that cross through more than one grids. Therefore, one road segment can belong to more than one grids.
Assume that the coordinate of the point at the top left corner is $P_0(x_0,y_0)$. When we need to locate a GPS location $P(x,y)$, we first calculate the grid it belongs to. Assume this grid is $g$. Then we check the road segments indexed in $g$ one by one to find the road segment that $P$ belongs to.

We can see that there is a tradeoff between grid size and storage cost. If the grid size is too small, it will take us too much space to store the information of index structure. And if gird size is relatively big, the number of roads we have to check seriatim also becomes big. In this circumstance, we propose a vehicle tracking based method to fulfill map-matching. Our key idea is that if the time difference between two successive locations of a vehicle is small, a vehicle can't run for long distance. Assume that the vehicle now locates at road segment $r$, with the help of outward neighbors (an outward neighbor of a road segment $r$ is a road segment that takes $r$'s exit as its own entrance), we can take the outward neighbors and recursively the outward neighbors of these outward neighbors as candidate road segments. Fig.\ref{fig-Map Matching}(b) shows an example of this strategy. The red point in Fig.\ref{fig-Map Matching}(b) is the last location of vehicle $v$, it locates on $r_4$, so the first class candidate roads for the next location of $v$ are the yellow ones, which are the direct outward neighbors of $r_4$, and the second class of that is the green ones which are the outward neighbors of the yellow ones.

\subsection{Travel Speed Calculation \& Vacancy Completion}\label{section-Travel Speed Whole}
Now we introduce the way we calculate travel speeds for the roads where vehicular data is available, measure spatial-temporal correlations and explain how to deal with travel speed vacancies.

\subsubsection{Travel Speed Calculation for Roads with Available Vehicular Data}\label{section-Current TS when data available}
\quad

At the end of current calculation interval, the accumulation of records for this interval is finished. So we can calculate travel speeds for road segments that are covered according to Definition \ref{defn-Covered}.
For later use, we define the network distance between two points $P_1$ and $P_2$ as,
\begin{defn}\label{defn-dist for points}
$dist(P_1,P_2)$ is the distance walking from $P_{1}$ to $P_{2}$ along the shortest lawful driving path between them.
\end{defn}

Then we define,
\begin{defn}\label{defn-central point}
The \emph{central point of a road segment} $r$, denoted as $cp(r)$, is the point that at the central location of $r$, when walking along $r$ from its entrance to exit.
\end{defn}
%
%
\begin{figure}[h]
  \centering
  \includegraphics[width=2.8in]{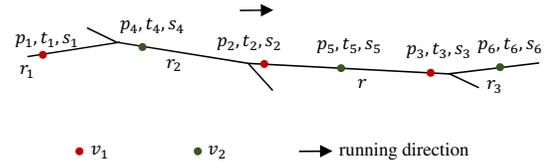}

  \caption{Road segments and vehicle locations.}
  \label{fig-Travel Speed With Available Data}
\end{figure}

We take the situation shown in Fig.\ref{fig-Travel Speed With Available Data} as an example to illustrate our calculation strategy. Assume the end time of current interval is $t_{e}$, so its begin time is $t_{b}=t_{e}-T$.
Suppose that vehicle $v_1$ runs from location $p_1$ to location $p_2$, then to $p_3$, and vehicle $v_2$ runs from $p_4$ to $p_5$, then to $p_6$. The value $s_i$ is the instant speed of the vehicle at its corresponding location. Assume that $t_1,t_4<t_{b}$, and $t_2,t_3,t_5,t_6\in [t_{b}, t_{e})$, and there is no other vehicle on $r$ during $[t_{b}, t_{e})$.

We then calculate the two-phase average speeds of $v_1$ from $p_1$ to $p_3$ as,
\[
\overline{s_{1_1}}=dist(p_1,p_2)/(t_2-t_1),
\overline{s_{1_2}}=dist(p_2,p_3)/(t_3-t_2).
\]
Similarly, the average speeds for $v_2$ from $p_4$ to $p_6$ are,
\[
\overline{s_{2_1}}=dist(p_4,p_5)/(t_5-t_4),
\overline{s_{2_2}}=dist(p_5,p_6)/(t_6-t_5).
\]

Then, we give the average speed for $r$ in this interval $T$ as,
\[
  S={(\overline{s_{1_1}}+\overline{s_{1_2}}+\overline{s_{2_1}}+\overline{s_{2_2}}+s_2+s_3+s_5+s_6)}/{(4+4)},
\]
where the first integer 4 in denominator is the number of average speeds calculated above, and the second 4 is that of speeds extracted from records that are right on $r$.

As demonstrated, we composite the instant speeds reported by vehicles and the average speeds calculated by locations and timestamps to generate the average speed for $r$. This composition leverages both explicit and implicit information of vehicular data.

\subsubsection{Cross Correlation Function}\label{section-sub-CCF}
\quad

The traffic features at upstream and downstream road segments are correlated. In our paper, this feature is travel speed. We use time-lagged cross correlation function to quantize this correlation.
Cross correlation function measures the inter-relevance between two vectors. We use Fig.\ref{fig-Travel Speed With Available Data} to explain our cross correlation calculation procedure.
In Fig.\ref{fig-Travel Speed With Available Data}, $r_1, r_2$ are upstream road segments to $r$, and $r_3$ is a downstream road segment from $r$.  Denote the travel speed of $r_1$ at $j^{th}$ calculation interval as $x_{j}$. Suppose that on average, vehicles on $r_1$ spend $k$ calculation intervals to run from the central point of $r_1$ to the central point of $r$. We call the number $k$ as the \emph{time lagging factor}. Then we denote $y_{j+k}$ as the travel speed of $r$ at $(j+k)^{th}$ calculation interval.
We use $X$ and $Y$ to represent the random variable forms of $x$ and $y$ respectively.
Then the cross correlation between $X$ and $Y$ is,
\begin{equation}\label{equ-CCF}
  c(X,Y)=\frac{{\gamma}_{XY}(k)}{\sigma_{X}\sigma_{Y}}, k=0,\pm1,\pm2,\pm3,\cdots
\end{equation}
where
$\gamma_{XY}(k)=E[(x_j-\mu_x)(y_{j+k}-\mu_y)]$,
with $\mu_x$ and $\mu_y$ being the means of $X$ and $Y$.
$\sigma_{X}, \sigma_{Y}$ are the  standard deviations of $X$ and $Y$.

\begin{figure}[h]
\centering
{
\includegraphics[width=\columnwidth]{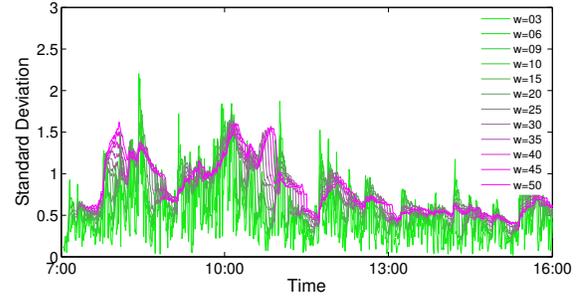}
}
\caption{The standard deviations of travel speeds for different length of $w$ when time advances. Here, we set $T=60s$. We can see that the standard deviations are relatively low. And these values first decrease and then increase when $w$ increases. }
\label{fig-OneCentralRoadsTS}
\end{figure}

A proper use of cross correlation function requires that the random variables $X$ and $Y$ satisfy stationarity condition. In travel speed's sense, it means that the travel speed of a road segment $r$ fluctuates around a constant value.
According to life experiences, we observe that for a road, its travel speed can't satisfy stationarity for long time as a consequence of the variation of traffic pattern. However, travel speed can satisfy it within a short period of time. We denote the length of slide window as $W$. It contains integral number of calculation intervals, which is denoted as $w=W/T$. More convincingly, we show our statistical result of the  travel speeds variation of a road in central city with different $w$ in Fig.\ref{fig-OneCentralRoadsTS}. It shows the stationarity of travel speeds.

\subsubsection{Vehicle Tracking Based Self-adaptive Time Lagging Factor Determination}\label{section-sub-Time Lagging}
\quad

In order to calculate the travel speed cross correlation between two road segments accurately, the time lagging factor $k$ between them must be determined properly.
We denote the time lagging factor between two road segments $u,r$ as $k_{u,r}$. Note that during different time of the day, $k_{u,r}$ also differs. For example, at morning peak, $k_{u,r}$  is bigger than that of midnight due to traffic congestion at morning. Therefore, it is necessary to determine $k_{u,r}$ for different time adaptively.

Novelly, we use the vehicle tracking to determine $k_{u,r}$. By recording the locations of a vehicle at different timestamps, we can obtain the time consumption for this vehicle to run from $u$ to $r$. Then by tracking a certain number of such vehicles, we can estimate $k_{u,r}$ in  average sense.
Specifically, we record the vehicles ever run on $r$ in the time duration of $W$. We put these vehicles in set $V_r$. As the upstream of $r$, there are some vehicles in $V_r$ that ran from $u$. By checking records, we select out vehicles satisfying this criterion and put them in set $V_{u,r}$. Note that the corresponding timestamps when vehicles in $V_{u,r}$ on $u$ may be prior to the beginning of $W$. For every vehicle $v$ in $V_{u,r}$, it has at least one record on $u$ and at least one on $r$. For $u$, we select the record whose location is nearest to the central point of $u$. If more than two of them are selected, we further select the record whose timestamp is the biggest. We call this selection strategy as $s^{'}(v,u)$. For $r$, we select the record whose location is nearest to the central point of $r$ while its timestamp is the smallest one in multiple records which satisfy this criterion. We call this selection as $s^{''}(v,r)$. With these two kinds of selection results of each $v\in V_{u,r}$, we then calculate average speed for $v$ from $u$ to $r$ as,
\[
avg(u,r,v)=\frac{dist(loc(s^{''}(v,r)),loc((s^{'}(v,u))))}{time(s^{''}(v,r))-time(s^{'}(v,u))}.
\]
Then we calculate the travel time of $v$ from $cp(u)$ to $cp(r)$ at the speed of $avg(u,r,v)$, namely,
\[
travelT(v)={dist(u,r)}/{avg(u,r,v)}.
\]
With the values of $travelT(v_i)$, we have the estimation of $k_{u,r}$ as,
\[
k_{u,r}=floor\left(\frac{\sum_{i=1}^{i=|V_{u,r}|}travelT(v_i)}{|V_{u,r}|*T}\right).
\]
where $floor(\cdot)$ gives the value of the floor of a decimal.

\begin{figure}[h]
  \centering
  \includegraphics[width=3in]{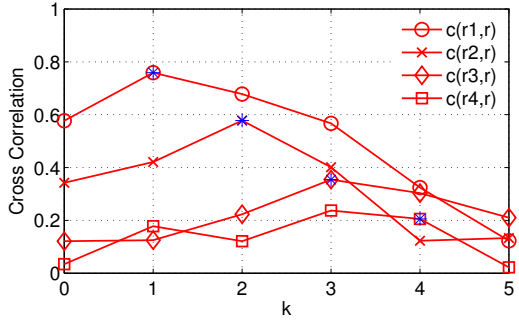}
  \caption{The comparison of the values of $k_{u,r}$ between our estimation method and the predefined values (\ie 0-5). These roads $r_1, r_2, r_3, r_4$ and $r$ are selected in central city of Shenzhen.  The values of $k$s corresponding to blue stars in this figure are the estimated time lagging factor values by our method. It can be seen that our method usually estimates the most proper $k$s that corresponding to highest cross correlation values. The value of $T$ is 60s, and the value of $w$ is set to 10. The time expansion of this figure is from 8:00:00 to 9:00:00 in April 18th 2011.}
  \label{fig-different time lagging factors for u and r}
\end{figure}

Therefore, through vehicle tracking, we get the estimation of the time lagging factor between two road segments during $W$.
To show the feasibility of this estimation method, we illustrate the cross correlation calculation results between two roads that are selected in central city for our $k_{u,r}$ and predefined time lagging factors in Fig.\ref{fig-different time lagging factors for u and r}. It shows that compared with predefined time lagging factors, the $k_{u,r}$ determined by our method gives more evident cross correlation values.

\begin{figure}[h]
  \centering
  \includegraphics[width=3in]{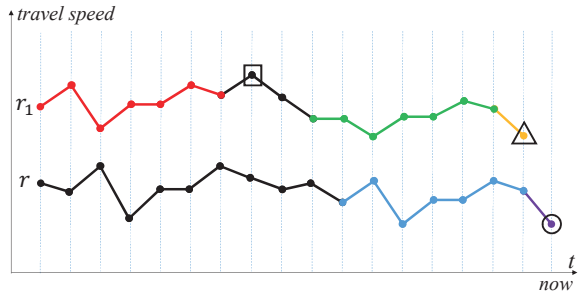}
  \caption{An explanation for the choice of $k$. The travel speed value in purple (surrounded by a circle) in Fig. \ref{fig-kNotMostCCF} needs to be estimated. Here $w=7$. So, if we search for the travel speed value sequence in $r_1$'s values which has the biggest cross correlation with the blue ones in $r$'s values, we will locate to the red sequence. Then the purple value will be estimated in reference of the value that surrounded by a rectangle. However, the rectangle one has increased from the last red one, while the purple one is lower than the last blue one. So their trends are not the same. Differently, though the green sequence has a lower cross correlation with the blue one, the value in yellow (surrounded by a triangle) decreases after the last green value, which is similar to the trend of the purple one.}
  \label{fig-kNotMostCCF}
\end{figure}

However, it is noteworthy that the most proper value of $k$  does not necessarily correspond to the highest value of cross correlation. Fig.\ref{fig-kNotMostCCF} illustrates the explanation for this. This is why we don't use a traversal way to find the lagging time corresponding to the highest cross correlation as the $k$ we need.

\subsubsection{Travel Speeds Estimation for Roads Where Vehicular Data Is Unavailable}\label{section-Instant Travel Speed Estimation}
\quad

\emph{(1) A Minimization Problem for Single-Road Vacancy}
\quad

In section \ref{section-Current TS when data available}, we showed our method of travel speed calculation for roads where data for current calculation interval is available. Now we handle single-road vacancy by using cross correlation function.
First, we define,
\begin{defn}
$X(index_1,index_2)$ is the sub-vector of vector $X$ from the subscript  $index_1$ to the subscript $index_2$.  And $X(index)$ is the item at the location $index$ which begins from 1. For example, if $X=[1,3,5,7,9]$, then $X(2,3)$ gives [3, 5], and $X(3)$ gives 5.
\end{defn}

Suppose that current calculation interval is the $n^{th}$ one from the beginning of $\mathcal{T}$, and now we want to fill the vacancy of the current travel speed for road $r$. Denote the area in which the roads we will consider to help to calculate the travel speed for $r$ as $A_r$, and denote the upstream roads of road $r$ within $A_r$ as $R_r=\{r_1, r_2, r_3, \cdots\}$. We then calculate the cross correlations between every $r_i$ and $r$.

\begin{figure}[!h]
  \centering
  \includegraphics[width=\columnwidth]{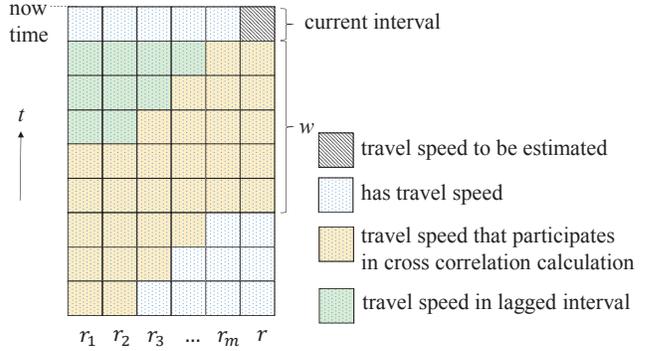}\\
  \caption{The interval selection to calculate $c_{pre}(r_i,r)$. Here $k_{r_1,r}=3$, $k_{r_2,r}=3$, $k_{r_3,r}=2$, $k_{r_m,r}=0$.}
  \label{fig-illustration of CCFPre}
\end{figure}

Denote the travel speed vector for $r_i$ from the $1st$ to the $n^{th}$ interval in $\mathcal{T}$ as $X_{r_i}$, and that for $r$ as $X_r$. As introduced in section \ref{section-sub-Time Lagging}, we can estimate the time lagging factor from $r_i$ to $r$. Then for any $r_i\in R_r$, the cross correlation between $r_i$ and $r$ during last $W$ which starts at $(n-w)^{th}$ and ends at $(n-1)^{th}$ calculation interval can be calculated as,
\begin{equation}\label{equ-cPre(ri,r)}
\begin{split}
& c_{pre}(r_i,r)=\\
& c(X_{r_i}(n-k_{r_i,r}-w,n-k_{r_i,r}-1),X_r(n-w,n-1)).
\end{split}
\end{equation}
Fig. \ref{fig-illustration of CCFPre} illustrates the interval selection for the calculation of $c_{pre}(r_i,r)$.

\begin{figure}[!h]
  \centering
  \includegraphics[width=3in]{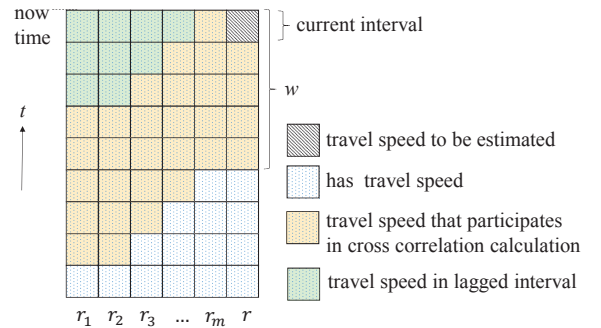}\\
  \caption{The  interval selection to calculate $c_{now}(r_i,r)$.}
  \label{fig-illustration of CCFNow}

\end{figure}

Similarly, the cross correlation between $r_i$ and $r$ during  $W$ which starts at $(n-w+1)^{th}$ and ends at $n^{th}$ calculation interval can be expressed as,
\begin{equation}\label{equ-cNow(ri,r)}
\begin{split}
& c_{now}(r_i,r)=\\
& c(X_{r_i}(n-k_{r_i,r}-w+1,n-k_{r_i,r}),X_r(n-w+1,n)).
\end{split}
\end{equation}
We illustrate the interval selection of the  calculation of $c_{now}(r_i,r) $ in Fig.\ref{fig-illustration of CCFNow}. It is noteworthy that $c_{now}(r_i,r)$ contains the unknown quantity $X_{r}(n)$.

Let vector $C_{pre}(r)=[c_{pre}(r_1,r),c_{pre}(r_2,r), \cdots]$ denote the values of $c_{pre}(r_i, r)$ between every $r_i$ and $r$.  And put the values of $c_{now}(r_i,r)$ between every $r_i$ and $r$ in vector $C_{now}(r)=[c_{now}(r_1,r),c_{now}(r_2,r),\cdots]$.

\begin{figure}[h]
\centering
\subfigure[The changes of cross correlations between different roads and $r$ when time advances. We can see that the cross correlation value between road pairs fluctuates in a small range.]
{
\includegraphics[width=\columnwidth]{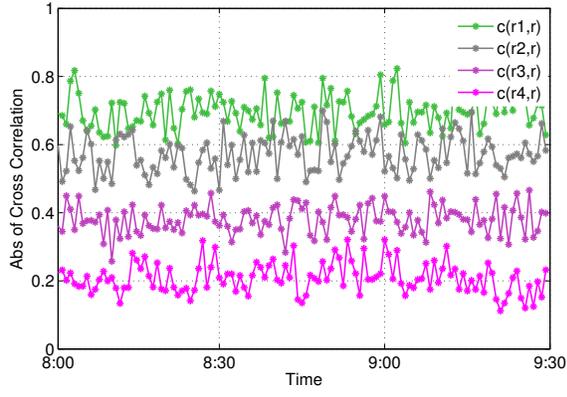}
}
\subfigure[The mean value and standard deviations of the cross correlations shown in Fig.\ref{fig-ccf-stationary}(a). It can be seen that the standard deviations are small, which indicates the stationarity of cross correlations.]
{
\includegraphics[width=\columnwidth]{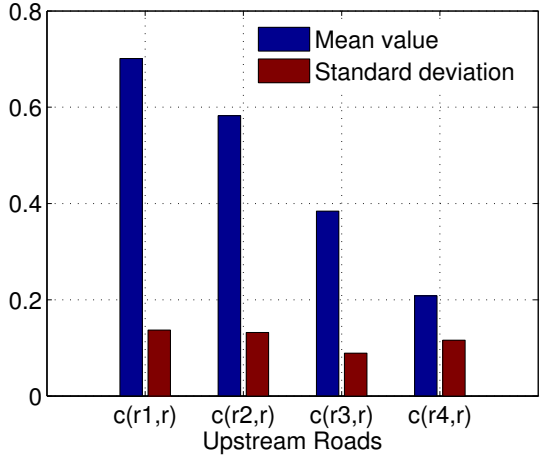}
}
\caption{The illustration of the stationarity of travel speeds cross correlation between roads.}
\label{fig-ccf-stationary}
\end{figure}
We observe that the cross correlation of travel speeds between roads also meet stationarity demand, which is shown in Fig.\ref{fig-ccf-stationary} as an example.
So we can estimate the value of the current travel speed $X_r(n)$ for $r$ by minimizing the differences between $C_{pre}(r)$ and $C_{now}(r)$, namely,
\begin{equation}\label{equ-min-obj}
obj=\mathop {\min} \limits_{X_{r}(n)} {||C_{now}(r)-C_{pre}(r)||}_{2},
\end{equation}
\ie
\[
obj=\mathop {\min} \limits_{X_{r}(n)} \left\{\sum_{i=1}^{m}\left[{c_{now}(r,r_i)-c_{pre}(r,r_i)}\right]^{2}\right\}^{\frac{1}{2}},
\]
where $m$ is the number of $r_i$s.
We then let
\[
f(X_r(n))=\sum_{i=1}^{m}\left[{c_{now}(r,r_i)-c_{pre}(r,r_i)}\right]^{2},
\]
so the minimization of $obj$ is equal to the minimization of $f(X_{r}(n))$.
For the simplicity of presentation, we let $Y_{r_i}=X_{r_i}(n-k_{r_i,r}-w+1,n-k_{r_i,r})$, and $Y_{r}=X_r(n-w+1,n)$, where $Y_{r}(w)=X_r(n)$.
Then $f$ can be further formulated as,
\[
f=\sum_{i=1}^{m}\left[{\frac{E[(y_r-\mu_{Y_r})(y_{r_i}-\mu_{Y_{r_i}})]}{\sigma_{Y_r}\sigma_{Y_{r_i}}}-c_{pre}(r,r_i)}\right]^{2}.
\]
We then let
\[
g_i(X_r(n))=E[(y_r-\mu_{Y_r})(y_{r_i}-\mu_{Y_{r_i}})],
\]
and
\[
h(X_r(n))=\sigma_{Y_r}\sigma_{Y_{r_i}},
\]
where $X_{r}(n)$ exists in $y_r$, $\mu_{Y_r}$ and $\sigma_{Y_r}$.
So $f$ can be shown in a simpler way as,
\[
f=\sum_{i=1}^{m}\left[{\frac{g_i}{h}-c_{pre}(r,r_i)}\right]^{2}.
\]
And the derivative of $f$ in terms of $X_{r}(n)$ is,
\begin{equation}\label{equation_f_deri}
f^{'}=2\sum_{i=1}^{m}\left[{\frac{g_i}{h}-c_{pre}(r,r_i)}\right]\frac{g_{i}^{'}h-g_ih^{'}}{h^{2}\sigma_{Y_{r_i}}}.
\end{equation}
Then by letting $f^{'}=0$, the value of $X_{r}(n)$ at which $obj$ reaches its minimum can be obtained.

\emph{(2) A Recursive Way for Global Vacancies Completion:}
\quad

Having the method for single road travel speed vacancy completion, we now fill all the vacancies of roads by using it in a recursive way.

As aforementioned, we use $A_r$ to limit the number of upstream road segments that help to calculate the current travel speed for $r$. More specifically, we use limited backtracking method to fulfill this aim. We set a distance threshold $d_A$ by combing network distance $dist()$ and the \emph{intersection distance}.
The intersection distance $distI(u,r)$ between road segment $u$ and $r$ is the number of intersections between them along the path of their network distance $dist(u,r)$. For example, $distI(r_1,r_3)$ in Fig.\ref{fig-Travel Speed With Available Data} is three.
We use the product of intersection distance and network distance to limit $A_r$, $Dist(u,r)=dist(u,r)\times distI(u,r)$. Only when $Dist(r_i,r)\leqslant d_A$, the upstream road segment $r_i$ should be taken into consideration to calculate the travel speed for $r$.

%

The use of cross correlation function is based on the assumption that the current interval travel speed values for every $r_i\in R_r$ are available, but in reality it is inevitable that some of them are also empty. Note that when traffic is under low pressure, the sparsity of vehicular data also becomes serious. So we define a threshold $N_{min}$. When the number of road segments having travel speeds contained in $R_r$, denoted as $N_{R_r}^{'}$, satisfies $N_{R_r}^{'} \geqslant N_{min}$, we stipulate that the calculation for $r$ can be continued, which means that we only use the roads having travel speeds in $R_r$ to calculate the travel speed for $r$. We only calculate the cross correlations between them and $r$. When $N_{R_r}^{'}$ satisfies above criterion, we say that $r$ is $calculable$.

Now, we use a recursive algorithm to fill global travel speed vacancies.
We start from a road segment $r$ which has no travel speed for current interval. Then we check each road segment $r_i \in R_r$ one by one from vicinity to remote.
If $r$ is calculable, we simply calculate its travel speed with the help of $R_r$. If $R_r$ contains any $r_i$ that does not have travel speed, we put them in $R_{r}^{'}$. For every $r_i \in R_{r}^{'}$, we take it as a road whose travel speed need to be filled one by one also from vicinity to remote, and check $R_{r_i}$ and $R_{r_i}^{'}$ for it. We show the whole algorithms for travel speeds completion in Alg.\ref{alg-current travel speeds filling-up for big R.} and Alg.\ref{alg-current travel speeds filling-up for r.}.

\begin{algorithm}[h]
\small
\caption{$TS\_All(\mathcal{R})$: Travel speeds calculation of $\mathcal{R}$.}
\label{alg-current travel speeds filling-up for big R.}
\begin{algorithmic}
\REQUIRE The edge set $\mathcal{R}$ of $\mathcal{G}$.
\ENSURE  The estimated travel speeds  for every $r \in \mathcal{R}$.
\FOR {each $r \in \mathcal{R}$}
\IF {$X_{r}(n)$ is not calculated}
\IF {$r$ has records for current interval}
 \STATE Calculate $X_{r}(n)$ using method introduced in section \ref{section-Current TS when data available};
 \ELSE
 \STATE call $TS(r)$;
 \ENDIF
\ENDIF
\ENDFOR
\end{algorithmic}
\end{algorithm}

\begin{algorithm}[h]
\small
\caption{$TS(r)$: Travel speed filling-up for $r$.}
\label{alg-current travel speeds filling-up for r.}
\begin{algorithmic}
\REQUIRE Road segment $r$.
\ENSURE  The estimated travel speed $X_{r}(n)$ for $r$.
\STATE stack $S\leftarrow \emptyset$;
\IF{$r$ is \emph{calculable}}
\STATE calculate $X_{r}(n)$;
\ELSE
\STATE $push(S, r) $;
\FOR {each $r_i \in R_{r}^{'}$}
\STATE call $TS(r_i)$;
\ENDFOR
\STATE calculate $X_{r}(n)$;
\STATE $pop(S, r)$;
\ENDIF
\end{algorithmic}
\end{algorithm}
\emph{(3) Initialization of Our Algorithm}
\quad

Note that our estimation method requires that at the $n^{th}$ interval, the travel speeds of past $w-1$ calculation intervals are all available for every $r \in \mathcal{R}$. So, when $n\leq w$, we have to fill up the vacancies in another way. We do this by averaging the history values. If no value was calculated in history, we set the value as a properly high speed because no data always means little number of vehicles running on road, which usually means a good traffic condition. This is the initialization procedure of our strategy.

\emph{(4) Parallelization of Our Algorithm}
\quad

With the purpose of accelerating the calculation speed of global travel speeds completion using our recursive algorithm, we can divide the whole area of city to several un-overlapped regions whose size are similar. Then in each of these regions, we choose a road as its starting location of recursive algorithm. By doing so, our method can be executed  in a parallelized way. The feasibility of this parallelization is guaranteed by the wide spatial distribution of vehicles.
\section{Experimental Evaluation}\label{section-Experimental Evaluation}
In this section, we illuminate the superiority of our travel speed calculation method by comparing its estimation errors with classical methods.
\subsection{Estimation Error Measurement}
Suppose that $\mathcal{S}=[s_1, s_2,\cdots ]$ contains the ground truth values of travel speeds for every road segment in $\mathcal{R}$, and $\mathcal{S}^{'}=[s_1^{'}, s_2^{'},\cdots]$ contains the estimated values of these travel speeds. The relative error of $\mathcal{S}^{'}$ to $\mathcal{S}$ is the relative difference between them, 
\[
 \varepsilon(\mathcal{S}, \mathcal{S}^{'})=\frac{{||\mathcal{S}^{'}-\mathcal{S}||}_{2}}{{||\mathcal{S}||}_{2}}=\frac{{(\sum_{i=1}^{|\mathcal{S}|}{|s_i^{'}-s_i|}^{2})}^{\frac{1}{2}}}{{(\sum_{i=1}^{|\mathcal{S}|}{s_i}^{2})}^{\frac{1}{2}}}.
\]
\subsection{Travel Speed Estimation Results}\label{section-expe-current}
As introduced, the calculation of estimation error needs ground truth. The real ground truth of travel speeds for roads is unavailable. So we use cross validation instead. We set the ground truth by picking up an experiment time duration in which the coverage ratio of roads is high. Then we randomly hide some travel speeds for some roads with a predefined missing ratio. Then we recover these missing(hidden) values by our completion strategy and comparative methods and then calculate their estimation errors respectively. Based on experimental analysis, the experiment duration we select is 7:00:00-16:00:00 April 24th, 2011, which averagely has 1512 out of 1882 roads covered by vehicles during every interval when $N_{thr}=2$. There are two reasons why we take 2 as the value of $N_{thr}$. Firstly, the sampling data collected by probe taxis is so coarse that the number of taxis running through a road segment is not big for a relative short interval, as we have shown in section \ref{section-sub-Problem of coverage incompletion in vecicular crowdsensing}. Secondly, if we take 1 as $N_{thr}$, it is not representative enough to implement our methods for the travel speed and time lagging factor calculation.

\begin{figure}[h]
  \centering
  \includegraphics[width=\columnwidth]{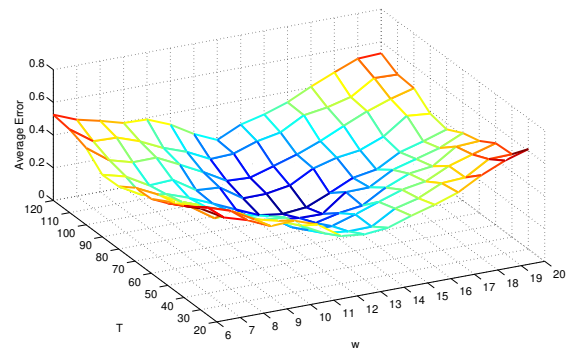}\\
  \caption{We randomly choose ten separated hours from the whole time expansion and calculate the average estimation error when the missing ratio is 0.2. Here, $d_A=2000m, N_{min}=4$.}
  \label{fig-fill-difft-w-STC}
\end{figure}

Before comparison, we should determine important parameters $T$ and $w$ because different combinations of their values give quite different estimation errors. We determine them by averaging the estimation errors corresponding to ten randomly chosen separated hours from the whole time expansion in data set. Fig.\ref{fig-fill-difft-w-STC} shows the average errors at different $T$ and $w$ when the missing ratio is 0.2. It can be seen that when $T=80s$ and $w=12$, the average estimation error is the lowest. Other missing ratios give similar suggestions.

Now we introduce the estimation methods that we will compare with.

$\bullet$\emph{Kriging} is a spatial interpolation method that builds variation function model according to the spatial location relation among data\cite{Ref82}\cite{Ref90}. Based on this variation function, Kriging interpolates the missing data using the correlation between the locations that have and not have data. It is a well-known data recover method but it mainly focuses on spatial relations of data at different locations.

$\bullet$\emph{KNN} uses the weighted average values of the K nearest locations to the location whose data to be estimated as the estimation value. Here we take the inverse of distance as weighting basis. The nearer is a location to the location to be estimated, the bigger its weight will be. And here we use 4 as the number $N$, as is always used in the literature.

$\bullet$\emph{ARIMA} (Autoregressive Integrated Moving Average Model) is a famous time series analysis based prediction method that regresses the dependent variable with regard to its lagging value and the present as well as lagged values of random error to establish models. It mainly considers the temporal relations among data \cite{Ref83}. Here we use the values in past $w-1$ intervals to estimate the values in $n^{th}$ interval.

\begin{figure*}[t]
  \centering
  \includegraphics[width=6.5in]{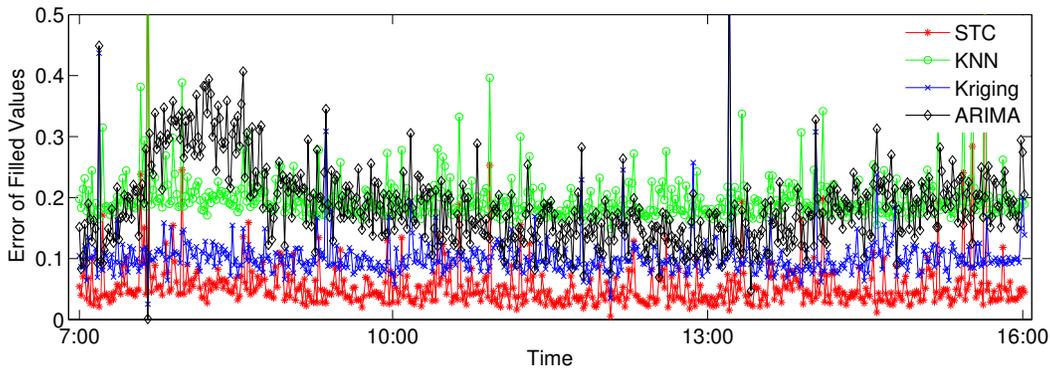}\\
  \caption{The comparison of filling-up errors among four estimation methods during 7:00-16:00 April 24th, 2011. The missing ratio is 0.2. And $T=80s, w=12$.}
  \label{fig-fill-compare-four-error-whole-time}
\end{figure*}

\begin{figure}[h]
  \centering
  \includegraphics[width=3in]{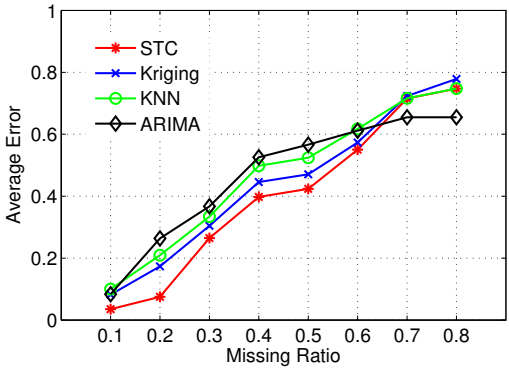}\\
  \caption{The average errors for different missing ratios when interval $T=80s, w=12$.}
  \label{fig-fill-compare-four-error-fill-average}
\end{figure}

Fig.\ref{fig-fill-compare-four-error-whole-time} shows the estimation errors of STC, Kriging, KNN, and ARIMA for every calculation interval in the experiment duration. The missing ratio in Fig.\ref{fig-fill-compare-four-error-whole-time} is set  to 0.2.  We can see that STC makes the lowest errors at most of the intervals.
And to observe the robustness of STC when the missing ratio varies, we show the average estimation error of travel speeds of all the intervals in the experimental duration in Fig.\ref{fig-fill-compare-four-error-fill-average}.  We can see that when the missing ratio is smaller than 0.7, STC always get the lowest errors among the four algorithms.

\subsection{An Extension for Future Travel Speed Prediction}
Besides current interval travel speed calculation, predicting future travel speed  also has great meaning for trip planning, transportation management, \etc. Our method can also be used for travel speed prediction for next calculation interval with a modification.
\subsubsection{Our Method for Predicting Future Travel Speeds}\label{section-Predicting Global Future Travel Speed}
\quad
Different from current interval travel speed estimation, the future travel speed prediction is based on the condition that all the values for current calculation interval is calculated but none of travel speeds for next interval is available. In this situation, we can predict travel values by a weighted averaging method.

Since no travel speed information in next interval is available, and the cross correlation function calculated only measures the correlation till current interval, to predict travel speed in next interval, we can only use the roads from which the lagging factor to $r$ is bigger than $0$. We pick up these road segments from $R$ and denote  the set containing them as $R_0$. For every $r_i\in R_0$, we have its corresponding $c_{now}(r_i,r)$. The value ${{c^2}_{now}(r_i,r)}$ indicates the \emph{coefficient of determination} that we can estimate the values of travel speeds for $r$ using the travel speed values of $r_i$. The weight we are about to use is,
\[
\omega(r_i,r)={c^2}_{now}(r_i,r)/\sum_{i=1}^{i=|R_0|}{{c^2}_{now}(r_i,r)}
\]

\begin{figure}[!h]
  \centering
  \includegraphics[width=\columnwidth]{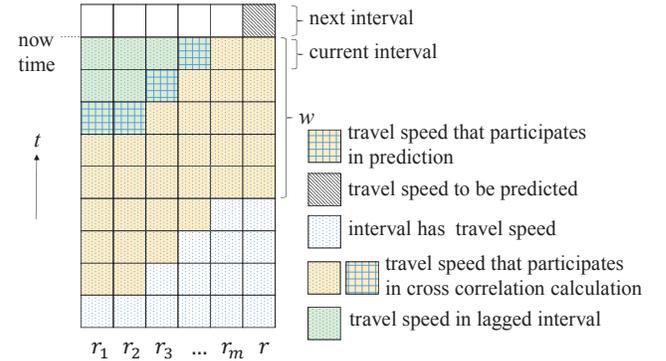}\\
  \caption{Interval selection for travel speed prediction.}
  \label{fig-illustration of Next}
\end{figure}

We then use linear regression to get the relationship between the travel speeds of $r_i$ and $r$. The regression result can be formulated as $X_{r}=a_{i}X_{r_i}+b_i$.
Therefore, the predicted travel speed of $r$ for next interval is,
\begin{equation}\label{equ-predict TS}
X_{r}^{+}(n+1)=\sum_{i=1}^{i=|R_0|}((a_i \times X_{r_i}(n-k_{r_i,r}+1)+b_i)\times \omega(r_i,r)).
\end{equation}
Fig.\ref{fig-illustration of Next} shows our interval selection for prediction.

\subsubsection{Future Travel Speeds Prediction Results}\label{section-expe-future}
\quad

\begin{figure}[h]
  \centering
  \includegraphics[width=\columnwidth]{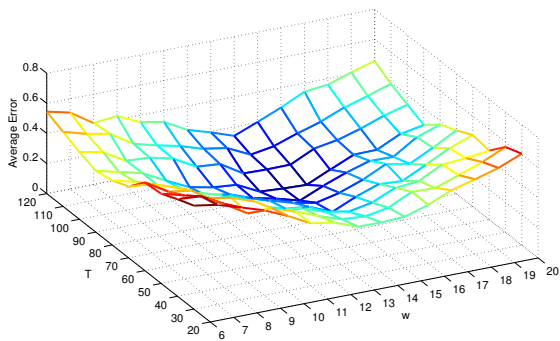}\\
  \caption{The average prediction errors of ten randomly chosen hours at different $T$ and $w$ when the missing ratio is 0.2.}
  \label{fig-predict-difft-w-STC}
\end{figure}

In future travel speeds prediction, we can use the next interval estimation results calculated using next interval vehicular data as ground truth. However, we still should determine the parameters $T$ and $w$ first. And similarly, we randomly choose ten separated hours in the experiment duration and change the values of $T$ and $w$. Fig.\ref{fig-predict-difft-w-STC} shows the results of this procedure. We can see that STC gets the lowest error when $T=90s$, and $w=13$.

Then we compare the the prediction error of STC with two  classical prediction methods:

$\bullet$\emph{KF(Kalman Filter)} is a theoretically optimal data processing algorithm\cite{Ref80}\cite{Ref92}\cite{Ref81}. It combines the values of last condition time and the current condition value and minimizes the mean of squared error. It can be used to predict system condition with favorable accuracy.

$\bullet$\emph{ARIMA} which was introduced before. In the literature, \eg Voort \etal \cite{Ref83} used ARIMA time series models to forecast traffic flow.

\begin{figure*}[t]
  \centering
  \includegraphics[width=6.5in]{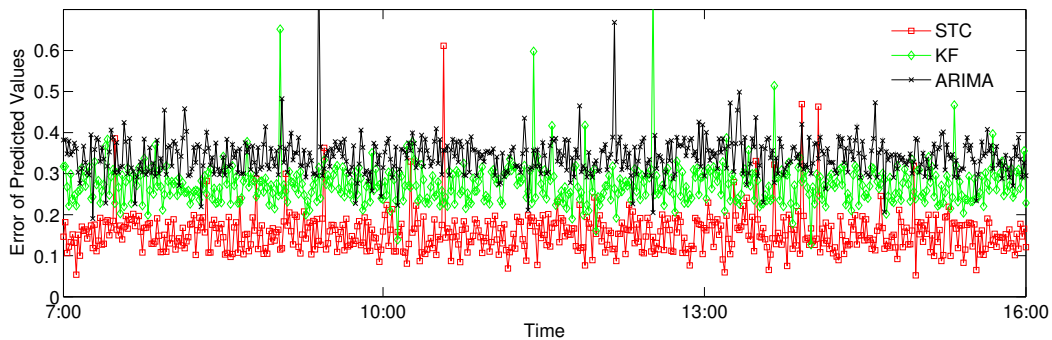}\\
  \caption{The prediction errors of STC, KF, and ARIMA. STC always gives the lowest errors among three methods.}
  \label{fig-predict-3-algs}
\end{figure*}

The comparison result is shown in Fig.\ref{fig-predict-3-algs} from which we can see that among three methods, our method always gives lowest errors in all the calculation intervals.

\section{Conclusion And Future Work}\label{section-Conclusion And Future Work}
In this paper, we propose a new vehicular crowdsensing data based travel speed calculation strategy called STC to estimate travel speeds. We use cross correlation to measure the spatial-temporal correlations of travel speeds among different road segments. When calculate cross correlation, we novelly use vehicle tracking to self-adaptively determine the time lagging factors of travel speed diffusion between roads. We settle the problem of single-road travel speed vacancy completion via reducing it to a minimization problem, using the local-stationarity of cross correlation. Then we fill up all the vacancies in a recursive way by utilizing the geometrical structure of road net. Experiments based on real vehicle data show that our strategy estimates travel speeds with lowest errors when compared with classical methods. And also, we show that our method can be easily transfered to predict travel speed for future calculation interval. In the future, we will pay more attention to the issues of wireless communication cost (\eg bandwidth consumption) saving and privacy protection in our design.

\bibliographystyle{IEEEtran}
\bibliography{references}

\end{document}